\documentclass[10pt,twocolumn,superscriptaddress,english,10pt,prl,showpacs,floatfix,aps]{revtex4-2}

\usepackage[latin9]{inputenc}
\usepackage{amsthm}
\usepackage{amsmath}
\usepackage{amssymb}
\usepackage{esint}
\usepackage{graphicx}
\usepackage{upgreek}
\usepackage{soul}
\usepackage{xspace}
\usepackage{braket}
\usepackage[backref=true,
bookmarksnumbered=true,
bookmarks=true,
bookmarksopen=true,
colorlinks=true,
citecolor=blue,
linkcolor=blue,
anchorcolor=green,
urlcolor=blue,unicode=false]{hyperref}

\makeatletter
\@ifundefined{textcolor}{}
{%
 \definecolor{BLACK}{gray}{0}
 \definecolor{WHITE}{gray}{1}
 \definecolor{RED}{rgb}{1,0,0}
 \definecolor{GREEN}{rgb}{0,1,0}
 \definecolor{BLUE}{rgb}{0,0,1}
 \definecolor{CYAN}{cmyk}{1,0,0,0}
 \definecolor{MAGENTA}{cmyk}{0,1,0,0}
 \definecolor{YELLOW}{cmyk}{0,0,1,0}
}

\usepackage{soul}
\usepackage{xspace}
\usepackage{braket}
\usepackage[backref=true,
bookmarksnumbered=true,
bookmarks=true,
bookmarksopen=true,
colorlinks=true,
citecolor=blue,
linkcolor=blue,
anchorcolor=green,
urlcolor=blue,unicode=false]{hyperref}

\renewcommand{\v}[1]{\ensuremath{\mathbf{#1}}} 
 
\let\baraccent=\= 
\renewcommand{\=}[1]{\stackrel{#1}{=}} 

\newcommand{\mos}{$\text{MoS}_\text{2}$}

\newcommand{\ws}{$\text{WS}_\text{2}~$}

\newcommand{\didv}{d$I$/d$V$}

\makeatother

\usepackage{babel}

\begin{document}

\title{Electronic and magnetic properties of single chalcogen vacancies in MoS$_2$/Au(111)}

\author{Sergey Trishin}
\affiliation{\mbox{Fachbereich Physik, Freie Universit\"at Berlin, 14195 Berlin, Germany}}

\author{Christian Lotze}
\affiliation{\mbox{Fachbereich Physik, Freie Universit\"at Berlin, 14195 Berlin, Germany}}

\author{Nils Krane}
\affiliation{\mbox{Fachbereich Physik, Freie Universit\"at Berlin, 14195 Berlin, Germany}}
\affiliation{\mbox{nanotech@surfaces Laboratory, Empa - Swiss Federal Laboratories for Materials Science and Technology, D\"ubendorf, Switzerland
}}

\author{Katharina J. Franke}
\affiliation{\mbox{Fachbereich Physik, Freie Universit\"at Berlin, 14195 Berlin, Germany}}

\begin{abstract}
Two-dimensional (2D) transition-metal dichalcogenides (TMDC) are considered highly promising platforms for next-generation optoelectronic devices. Owing to its atomically thin structure, device performance is strongly impacted by a minute amount of defects. Although defects are usually considered to be disturbing, defect engineering has become an important strategy to control and design new properties of 2D materials. Here, we produce single S vacancies in a monolayer of MoS$_2$ on Au(111). Using a combination of scanning tunneling and atomic force microscopy, we show that these defects are negatively charged and give rise to a Kondo resonance, revealing the presence of an unpaired electron spin exchange coupled to the metal substrate. The strength of the exchange coupling depends on the density of states at the Fermi level, which is modulated by the moir\'e structure of the \mos\ lattice and the Au(111) substrate. In the absence of direct hybridization of \mos\ with the metal substrate, the S vacancy remains charge-neutral. Our results suggest that defect engineering may be used to induce and tune magnetic properties of otherwise non-magnetic materials.

\end{abstract}

\maketitle

\section{Introduction}
Monolayers of semiconducting transition-metal dichalcogenides (TMDCs) and related two-dimensional (2D) materials have recently attracted a lot of attention due to their intriguing optoelectronic properties \cite{Wang2012}. The immense progress in flexible design of heterostructures has enabled the realization of device structures with unprecedented potential for efficient optoelectronic components \cite{Wang2012, Novoselov2016}. 
However, device performance and reliability remain still to be improved. Due to the 2D nature, already a minute amount of defects has an immense influence on the properties of the material. As material growth and processing usually come along with a variety of defects, it is of utmost importance to understand their effects \cite{Hu2018, Liang2021}. Most defects create states within the semiconducting band gap, and thus strongly modify the conductance and photoluminescence properties \cite{Qiu2013,Yu2014, KC2014,Nan2014,Chow2015,Lin2018,Chee2020,Yang2020,Bussolotti2021,Mitterreiter2021,Lin2022}, but also affect the chemical reactivity \cite{Li2016}. 

The most prevalent structural defects in TMDCs are chalcogen vacancies owing to their low formation energy \cite{Zhou2013,KC2014, Komsa2015, Lin2016,Tumino2020, Tan2020}. Similar to many defects, they introduce states in the pristine band gap, with their occupation, i.e. their charge state, being strongly influenced by the electrostatic environment \cite{Yu2014, Yang2019, Tan2020,Bussolotti2021, Akkoush2023}. 
Consequently, by adjusting the chemical potential of the monolayer's support, one can effectively control the charge state of the defects. When the defect state is singly occupied, the localized charge is accompanied by a local spin density. As a result, the TMDC material may exhibit paramagnetic or ferromagnetic behavior at high defect densities \cite{Li2018}. At very large defect densities or grain boundaries, the states may even interact and lead to the emergence of correlated states \cite{Tongay2012, Jolie2019, vanefferen2023}.

Here, we chose to investigate chalcogen vacancies in a monolayer of molybdenum disulfide (\mos), which is in contact to a Au(111) substrate. \mos\ belongs to the class of semiconducting TMDCs with an indirect bandgap in the bulk, which evolves into a direct bandgap in the single layer \cite{Splendiani2010, Mak2010}. The intact monolayer is inert as all Mo and S bonds are fully saturated, albeit with a strong interface hybridization at the Au--S interface \cite{Bruix2016,Tumino2020}. The most abundant point defects in this system are S vacancies at the interface to the Au substrate \cite{Tumino2020}. To systematically study S vacancies in the top layer, we deliberately bombard the sample with Ne ions. We show that these defects are charged and host a localized electron spin which is manifested in a Kondo resonance. When the \mos\ layer is not directly hybridized with the Au(111) substrate, there is no Kondo resonance and the states inside the semiconducting band gap are similar to those expected for S vacancies on free-standing \mos. Our results thus highlight the role of a metal support for controlling the charge state and magnetic properties.

\section{Experimental details and sample preparation}

We grew monolayer-islands of \mos\ on a pristine Au(111) substrate, which had been cleaned by repeated cycles of sputtering and annealing. We first deposited Mo atoms on the surface and subsequently annealed the sample at 800~K in H$_2$S gas at a pressure of $p=10^{-5}$~mbar. Inspection by STM assured the successful formation of \mos\ islands, which were characterized in more detail in earlier studies \cite{Krane2018a}. 
To introduce S vacancies in the top layer, we then irradiated the as-prepared sample with a beam of Ne ions with kinetic energy of 100~eV and an incidence angle of 55$^\circ$ with respect to the surface normal. 
 
The same preparation conditions were used in two different ultra-high vacuum chambers, connected to different STM set-ups. The first enabled the simultaneous investigation by atomic force microscopy (AFM) using a qPlus tuning fork at a temperature 4.5~K. The second set-up was used for temperature-dependent tunneling spectroscopy in a range spanning from 1~K to 28~K. Additionally, the sample could be subject to a magnetic field up to 3~T.

\section{Identification of S vacancies in \mos\ islands}

\begin{figure}
  \centering
  \includegraphics[width=0.95\columnwidth]{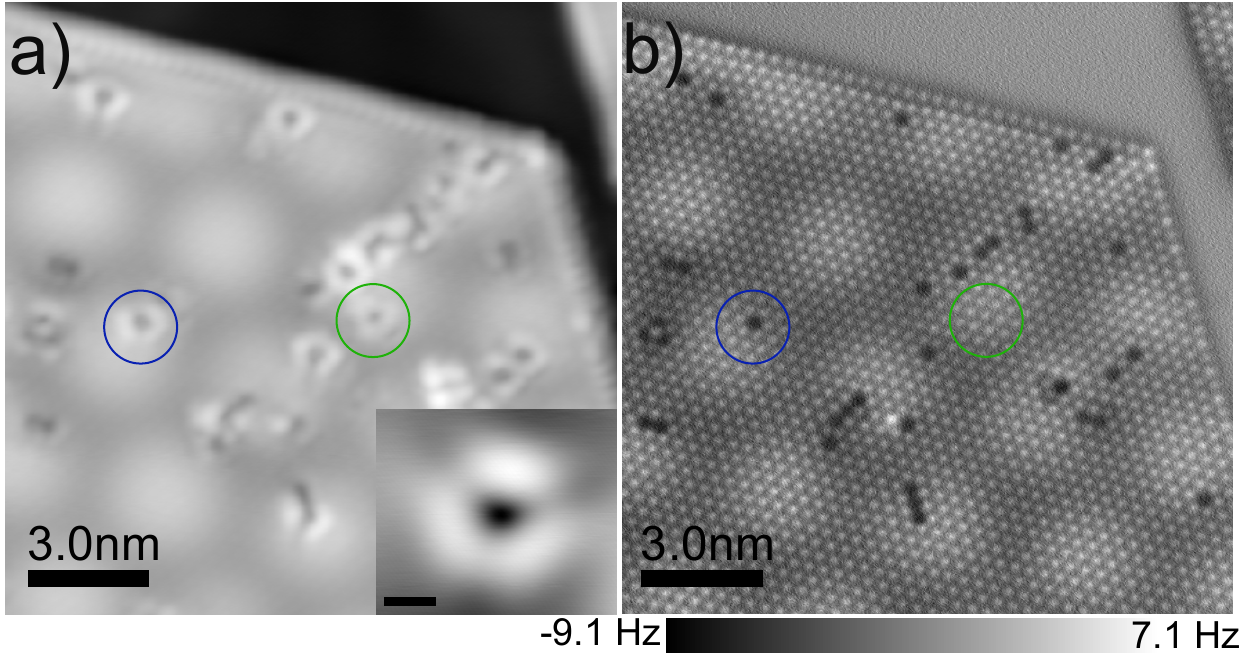}
  \caption{ Characterization of single top-layer sulfur vacancies in \mos. a) STM topography of a \mos\ island after sputtering. Several defects appear with a dark center and a bright threefold rim (close-up view in inset). The constant-current STM image was recorded at a setpoint of 100~mV and 200~pA, the inset at 300~mV and 200~pA. The scale bar in the inset is 5~\AA. b) Constant-height $\Delta\text{f}$ image, resolving the atomic structure of the terminating layer. Some S atoms are missing. These are at the location of the  bright-rim defects in (a).  The $\Delta\text{f}$ image was recorded at a setpoint of 1~mV and 200~pA.}
  \label{Sdef:1}
\end{figure}

We start by characterizing the structure of the \mos\ islands after Ne bombardment. The STM images (Fig.~\ref{Sdef:1}a) reveal the typical hexagonal structure originating in the moir\'e superstructure forming as a consequence of the lattice mismatch between the terminating S layer and the Au(111) substrate. At the same time, the image also reveals randomly distributed features, consisting of a dark center and a bright rim. A close-up view on one of these defects is displayed in the inset of Fig.~\ref{Sdef:1}a. The rim appears with a three-fold symmetric shape. A similar shape has been reported from theoretical and experimental studies of chalcogen vacancies in (other) TMDCs \cite{Gonzalez2016,Vancso2016,Schuler2019b,Tumino2020}. It is thus tempting to assign these defects to missing S atoms. Yet, to obtain complementary insights into the origin of the defects, we probed the frequency-shift ($\Delta\text{f}$) signal of the very same area (Fig.~\ref{Sdef:1}b). This image resolves the atomic structure of the top S layer, and confirms that some S atoms are missing. Comparison of the STM and AFM image allows us to associate most of the dark features surrounded by a bright rim to S vacancies. One of these is highlighted with the blue circle in Fig.~\ref{Sdef:1}a,b. When several S atoms are missing in neighboring sites, the rim shape is distorted. However, some few bright features in the STM image can not be assigned to missing top-layer S atoms as the top layer remains intact (green circles in Fig.~\ref{Sdef:1}a,b). Hence, these must originate in interstitials or bottom-layer sulfur vacancies \cite{Gonzalez2016} that have been incorporated in the layer during its growth process. 

\section{Electronic and magnetic properties of sulfur vacancies}
\label{sulfur:elec}

\begin{figure}
  \centering
  \includegraphics[width=0.95\columnwidth]{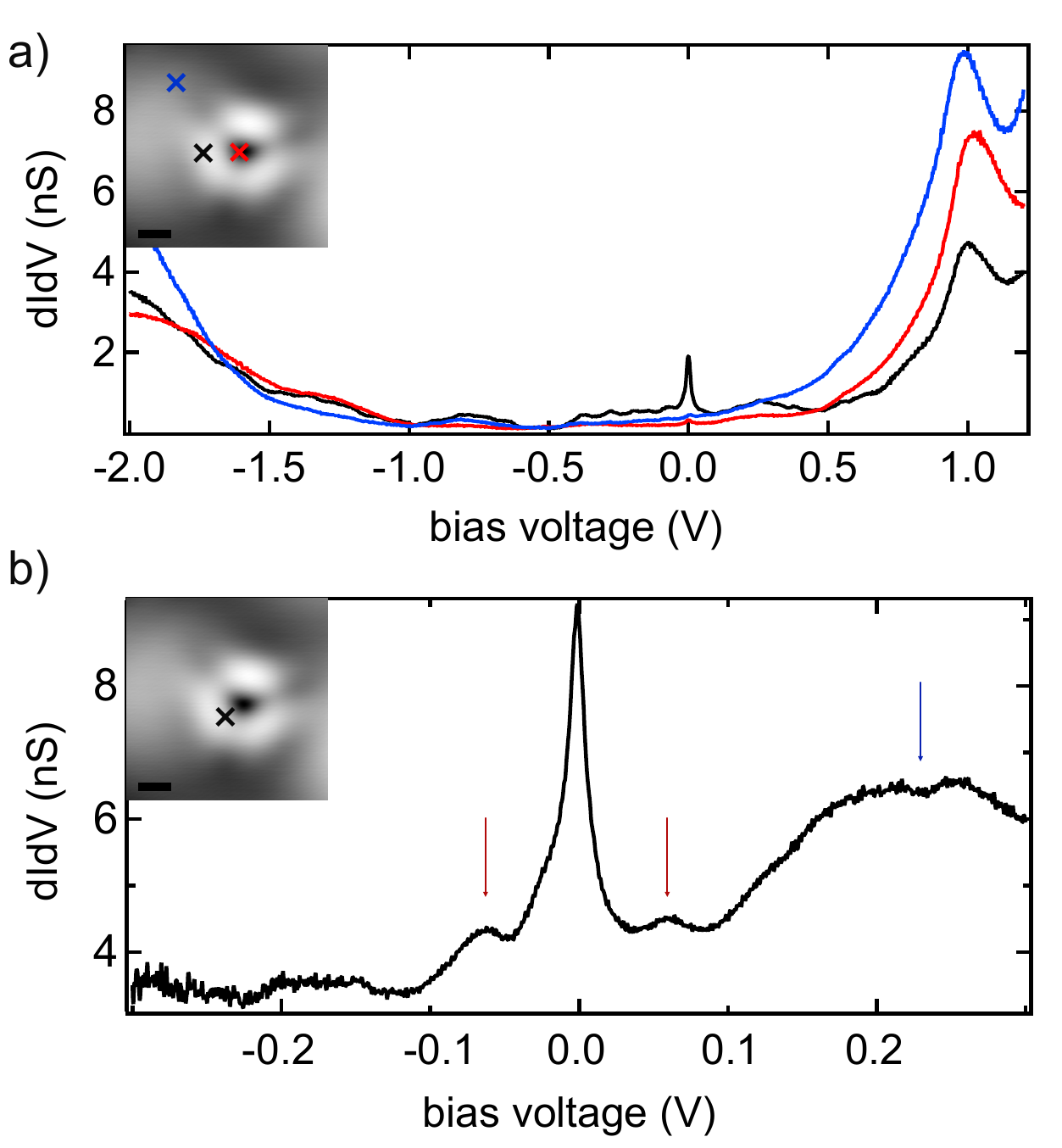}
  \caption{ Electronic structure of top-layer vacancies. a) \didv\ spectra recorded on the locations indicated in the insets, revealing the semiconducting band gap of \mos\ and a zero-bias resonance on the defect's rim (feedback opened at -2~V and 200~pA, $\text{V}_{\text{mod}}=\text{5}~\text{mV}$) . b) \didv\ spectrum around the Fermi level recorded on the rim of the defect (feedback opened at 300 mV and 200 pA, with an additional tip approach by 100 pm, $\text{V}_{\text{mod}}=\text{2}~\text{mV}$). The STM topographies in the insets were recorded at 300~mV and 200~pA. The scale bars in the insets are 5~\AA. }
  \label{Sdef:2}
\end{figure}

Having successfully created top-layer S vacancies, we now turn to their electronic characterization. For comparison, we first present a \didv\ spectrum on the intact \mos. It shows the semiconducting band gap, flanked by the onset of the valence band at -1.6~V and the onset of the conduction band at 0.9~V [Fig.~\ref{Sdef:2}a (blue)] \cite{Krane2018a}. A \didv\ spectrum recorded in the center of the S vacancy (red) looks very similar to the bare \mos\ spectrum, yet, with the onset and the peak of the conduction band being shifted by a few tens of mV. The most notable feature associated to the S vacancy is, however, found on the three-fold symmetric rim. Here, we detect a peak at the Fermi level, which is better brought out in the close-up view in Fig.~\ref{Sdef:2}b. Additionally, we observe two narrow resonances at $\pm$~60~mV and a broad one at $\approx$~230~mV.   

\begin{figure}
  \centering
  \includegraphics[width=0.95\columnwidth]{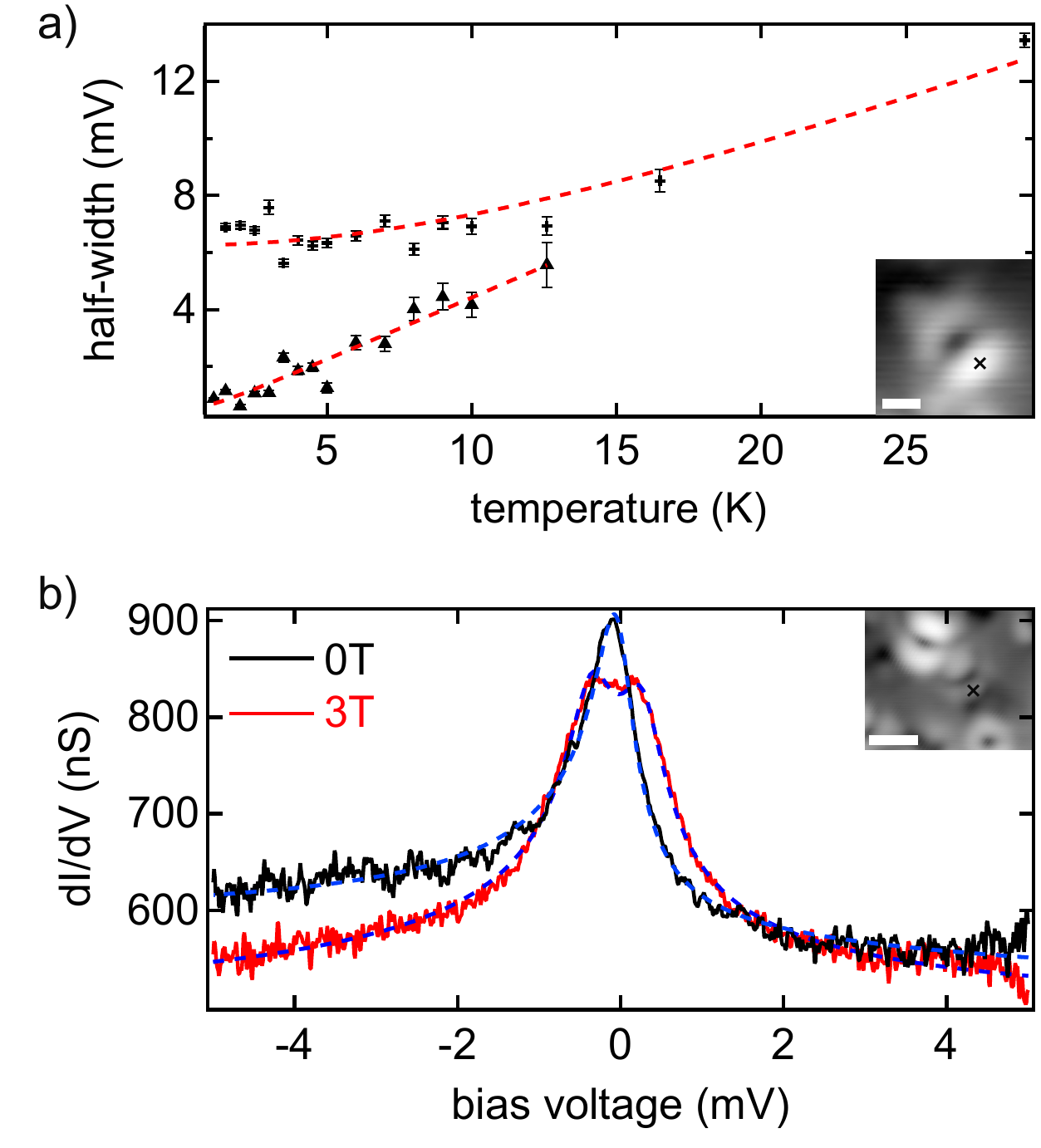}
  \caption{ Identification of the zero-bias peak as a Kondo resonance. a) Extracted half-widths of two vacancies [the topography of the one with the higher Kondo temperature is shown in the inset in (a), the other one in the inset in (b)], probed at different sample temperatures. To obtain the half-widths, we fitted the resonances with a Frota function and converted the Frota parameter to the half-width $\Gamma$ \cite{Prueser2011, Frank2015}. The Fermi-Dirac temperature broadening of the tip was subtracted from the line widths \cite{Goldhaber1998}. The temperature-dependent curve was fitted with the Fermi-liquid description of the Kondo problem \cite{Nozieres1974, Madhavan1998}. b) \didv\ spectra recorded on the rim of a vacancy indicated in the inset at zero magnetic field (black) and in an applied magnetic field of 3~T. The blue dashed lines represent fits with one (0~T) or two separated (3~T) Frota functions, which are split by $600\pm 100 \mu$V. The STM topographies in the insets were recorded at a setpoint of 50~mV and 3~nA (a) and 1 mV and 100 pA (b). The scale bars in the insets are 0.5~nm (a) and 1~nm (b). The \didv\ spectra were recorded after openning the feedback loop at 5~mV and 3~nA and with a modulation amplitude of $\text{V}_{\text{mod}}=\text{50}~\text{\textmu V}$.}
  \label{Sdef:3}
\end{figure}

To unravel the origin of the zero-bias peak, we investigated the evolution of \didv\ spectra of several S vacancies with increasing temperature. We fit all lineshapes by a Frota function (for an example see Fig.\ref{Sdef:3}b) and extract their half-widths $\Gamma$ after subtraction of the Fermi-Dirac temperature broadening \cite{Prueser2011, Frank2015}. The results are plotted in Fig.~\ref{Sdef:3}a for two distinct defects. Both peak broadenings can be reproduced by the phenomenological expression for the temperature dependence of a spin-1/2 Kondo resonance \cite{Nozieres1974, Madhavan1998,Nagaoka2002}:

\begin{equation}
\Gamma(T)=\sqrt{(\alpha k_{B}T)^{2}+2(k_{B}T_{K})^{2}},
\label{tkfit2}
\end{equation}

albeit with different Kondo temperatures of $T_{K}= 51\pm 2$~K and $T_{K}= 4\pm 3$~K, respectively. The temperature dependencies thus indicate the presence of single-electron spins around the defect sites, which are exchange coupled to the substrate. We comment on the origin of the different Kondo temperatures below.

To further support the interpretation of the zero-bias peak as a Kondo resonance, we probe its behavior in an external magnetic field of 3~T perpendicular to the sample. Figure ~\ref{Sdef:3}b shows such a \didv\ spectrum recorded on a S vacancy with a low Kondo temperature. The splitting of the resonance is in agreement with a Zeeman split of a spin-1/2 resonance and a $g$-factor of 2 (Fig.~\ref{Sdef:3}b). (Zero-bias peaks with significantly larger width, and thus larger Kondo temperatures, cannot be split in a magnetic field up to 3~T.) 
Following the assignment of the zero-bias peak as a Kondo resonance, the two narrow resonances at $\pm$~60~mV in the \didv\ spectra (Fig.~\ref{Sdef:2}b) can tentatively be assigned to excitation of the singly-occupied resonance at negative bias voltage, separated by the Coulomb charging energy $U=120$~mV to the doubly occupied state at positive bias voltage \cite{Goldhaber1998,Torrente2008}.

\begin{figure}
  \centering
  \includegraphics[width=0.95\columnwidth]{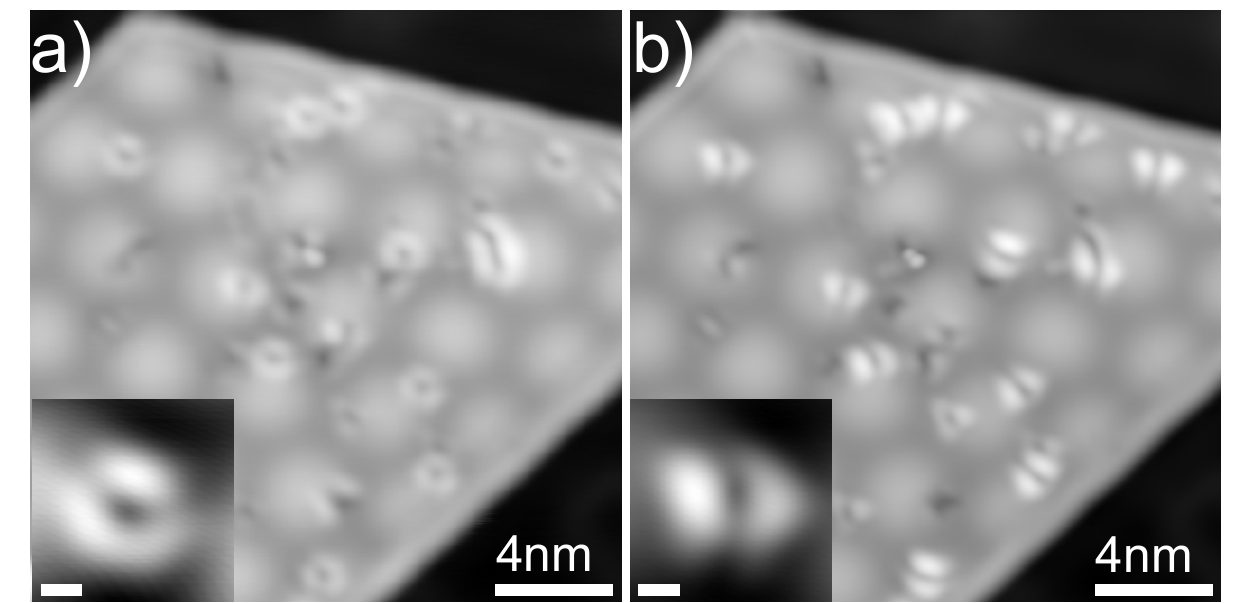}
  \caption{Influence of the moir\'e structure on the defect appearance.  a) STM images recorded at 300~mV, 200~pA. b) Same area and defect as in (a) recorded at 5~mV, 200~pA. Whereas the high bias voltage reveals a threefold shape, the images recorded close to the Kondo resonance show a broken symmetry, imposed by the moir\'e modulated density of states.}
  \label{Sdef:4}
\end{figure}

As already mentioned above, we find different Kondo temperatures for different S vacancies. The extracted Kondo temperatures vary in a range from 4-220~K. We ascribe such large variation to the different sites of the defects with respect to the moir\'e structure of the \mos\ layer. A similar observation was made for Fe and Mn atoms and was explained by the modulated density of states at the Fermi level by the moir\'e structure \cite{Trishin2021, Trishin2023}. 
Another indication of the moir\'e structure influencing the Kondo effect is found in the spatial distribution of the Kondo amplitude. While the electronic structure imaged at larger bias voltage revealed a three-fold symmetric rim (Fig.~\ref{Sdef:1}a and \ref{Sdef:4}a), the Kondo amplitude of most vacancies shows an almost two-fold symmetric shape (Fig.~\ref{Sdef:4}b and inset). These distributions reflect the local symmetry of the moir\'e structure with the mirror axis aligning along the valleys of the moir\'e modulation. Hence, all our observations point to a strong impact of the electronic modulation of the density of states on the Kondo effect. However, we cannot exclude that some of the S defects may be occupied by rest-gas molecules from the UHV chamber, such as oxygen, which are hard to be distinguished from the bare S vacancies \cite{Barja2019, Cochrane2021}. 

\begin{figure}
  \centering
  \includegraphics[width=0.95\columnwidth]{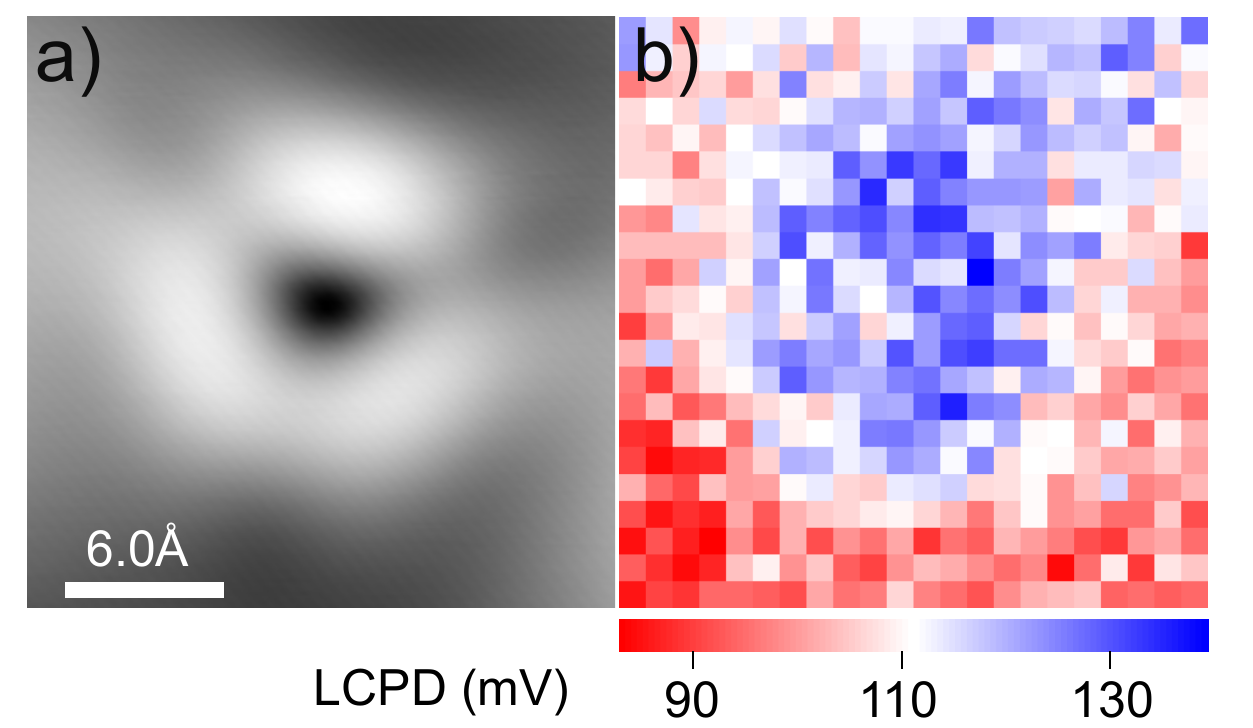}
  \caption{ Charge distribution around a sulfur vacancy. a) Close-up STM topography of a sulfur vacancy (recorded at 300~mV and 200~pA). b) LCPD extracted from a densely spaced grid of $\Delta\text{f}-\text{V}$ spectra over the same area as in (a). The $\Delta\text{f}-\text{V}$ spectra were recorded at a setpoint of 300~mV and 200~pA with the tip being additionally approached towards the surface by 100~pm.}
  \label{Sdef:5}
\end{figure}

An unpaired electron spin must originate in a localized charge. 
To investigate the local charge distribution we measure the local-contact-potential difference (LCPD) using a tuning-fork AFM. We extract the LCPD values from a grid of $\Delta\text{f}-\text{V}$ spectra (Fig.~\ref{Sdef:5}b). Comparing to the simultaneously recorded STM image, we associate the increase in LCPD to the S-vacancy site. A more positive LCPD value implies a negative charge localization, suggesting electron transfer from the metal to the defect site. This charge transfer is consistent with density-functional-theory calculations of S vacancies in \mos\ on Au(111)  \cite{Akkoush2023}. Concomitant to the charge transfer, theory also predicts a Jahn-Teller distortion of the atomic structure \cite{Tan2020, Akkoush2023}. However, the expected atomic displacements are very small ($\approx 0.3$~\AA) and not unambiguously detectable in our AFM images.

\section{Sulfur vacancies in patches of quasi-free-standing \mos}
\label{spit}

\begin{figure}
  \centering
  \includegraphics[width=0.95\columnwidth]{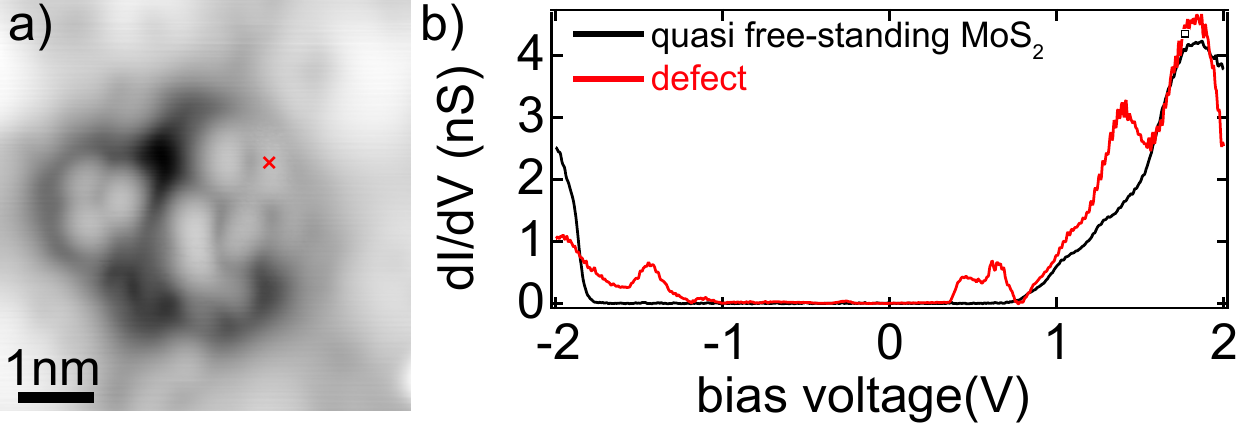}
  \caption{ Sulfur vacancies on quasi free-standing \mos. a) STM topography of point defects on a quasi free-standing \mos. b) \didv\ spectra recorded on a patch of quasi free-standing \mos (black) and on a point defect, the latter marked by a red cross in the topography in (a). The topography was taken at a setpoint of 800~mV and 100~pA, the spectra were recorded at a setpoint of 2~V and 3~nA (red) and 800~mV and 100~pA (black). }
  \label{Sdef:6}
\end{figure}

We have thus established that the Au substrate is of crucial importance for the formation of singly-charged S vacancies, which are Kondo screened. To obtain direct insight into the role of the metal, it would be illuminating to investigate the same vacancies in the absence of the metal substrate. To do so, we take advantage of another type of defect of the \mos/Au(111) sample. Some areas of the \mos\ islands, typically of several nanometers in diameter, appear 
slightly darker at large bias voltages and completely transparent at low bias voltage. These areal defects have been ascribed to the absence of a single atomic layer of Au atoms below the \mos\ \cite{Krane2016}. Figure~\ref{Sdef:6}a shows such an area (dark background) decorated by three S vacancies. 
Differential conductance spectra on such an area without S vacancies exhibit a band gap of similar width as expected for an isolated \mos\ monolayer \cite{Qiu2015}. The conductance in the gap is almost zero and the onsets of the valence and conduction band are much sharper than on Au (Fig.~\ref{Sdef:6}b). Hence, hybridization with the metal is negligible and we thus refer to these areas as quasi free-standing \mos\ monolayers \cite{Krane2016}. Spectra taken on the S-vacancy sites show a resonance at -1.4~V in the occupied states and two unoccupied states at 450~mV and 650~mV within this band gap, whereas the region close to the Fermi level is essentially flat. 
A state close to the valence-band edge along with a spin-orbit split state in the unoccupied regime of the band gap have been predicted for free-standing \mos\ and to arise from dangling bonds with strong Mo-$4d$ character \cite{Gonzalez2016,Tan2020, Sorkin2022, Akkoush2023}. Our observations are thus in agreement with a neutral vacancy state, which is stable in the absence of hybridization with a metal support, as e.g. also observed in WS$_2$ on silicon carbide \cite{Schuler2019b}. 

\section{Conclusions}

The characterization of defects in monolayers of TMDCs is of crucial importance for understanding the performance of electronic devices made of TMDC heterostructures. Here, we deliberately created S vacancies in monolayers of \mos\ grown on a Au(111) surface by Ne-ion bombardment. We showed that these defects are singly charged, which we ascribe to charge transfer from the metal substrate. The localized electron is Kondo screened by the electron bath at the interface of \mos\ to Au(111). Owing to a moir\'e-induced modulation of the density of states at the Fermi level, the exchange coupling to the substrate varies depending on the specific location of the S vacancy. Removal of a single Au layer in a small area below the \mos\ immediately suppresses charge transfer. 
Our results highlight the impact of a metal substrate for controlling the charge state and magnetic properties of single-atom defects in \mos.

\acknowledgments
We thank Blanca Biel, La\"etitia Farinacci, Mariana Rossi and Rika Simon for fruitful discussions. We acknowledge financial support by the Deutsche Forschungsgemeinschaft (DFG, German Research Foundation) through project number 328545488 (TRR 227, project B05).

%

\end{document}